\pdfoutput=1
\documentclass[final,5p,times,twocolumn]{elsarticle}

\usepackage{graphicx}

\usepackage{amssymb}

\journal{Nuclear Instruments and Methods A}

\begin{document}

\begin{frontmatter}

\title{Silicon Photomultipliers for High Energy Physics Detectors}


\author[A]{Erika Garutti}
\address[A]{Deutsches Elektronen-Synchrotron (DESY), Hamburg, Germany}

\begin{abstract}

In this paper I want to review the current status of development of multi-pixel silicon-based avalanche photo-diodes operated in Geiger mode, also known as Silicon Photomultipliers (SiPM). Particular emphasis is given to the application of this type of photo-sensors in high energy physics detectors. 

\end{abstract}


\end{frontmatter}

\section{Introduction}
Multi-pixel silicon-based avalanche photo-diodes operated in Geiger mode (SiPM) have become increasingly popular in HEP, as well as medical and astroparticle physics applications.  What makes them attractive alternatives to the more conventional photomultiplier tubes (PMT) is for instance the insensitivity to magnetic fields and in some cases the higher photon detection efficiency, the compactness and robustness, and the low cost. With respect to avalanche photo-diodes (APD) SiPMs offer a high intrinsic gain (10$^5$-10$^7$) which significantly simplifies the readout electronics, and increases the response stability to voltage or temperature changes. SiPMs are also insensitive to ionizing radiation,  i.e. the nuclear counter effect.
Additionally, SiPMs provide fast timing ($\sim$100\,ps FWHM for a single photon) and single photon counting capabilities. 
In this paper I review the products available on the market in terms of their basic properties and I discuss how each property influences or limits the design of typical HEP detectors. \\
The zoo of names referring to multi-pixel Geiger-mode avalanche photo-diodes (SiPM, G-APD, SSPM, MRS APD, AMPD, MPPC, etc.) is only partially representative of the increasing variety in performance between the various products. We shall see what the key parameters are to characterize a SiPM\footnote{I will adopt in this paper the name SiPM to generically refer to them all.} and how the various products are comparing on the basis of these parameters.

Good SiPMs by themselves do not make up for a whole HEP detector. Therefore, it is important to review the current status of SiPM specific readout chips as well as the implications of SiPMs on the design of a multi-channel detector. Issues like the need of individual SiPM bias adjustment and of the calibration and monitoring system are addressed. 

\section{SiPM basic properties}

The equivalent electric circuit of a SiPM is the parallel sum of as many diodes as the SiPM has individual pixels, typically 100-2000\,mm$^{-2}$. To all diodes a common inverse bias voltage is provided to operate the device a few volts above the breakdown point, $U_{bd}$. Each diode requires an active quenching of the Geiger discharge, normally obtained by connecting a resistor in series with the diode. 

\subsection{ Signal time characteristics} 

The time needed for a pixel to become active again after a discharge is known as the pixel recovery time, $\tau$. This is the product of the pixel quenching resistor, R$_q$, and the pixel capacitance, C$_{pix}$. The value of R$_q$ has to be larger than 100~k$\Omega$ to be able to quench the avalanche and can be as large as 20~M$\Omega$ in some cases~\cite{Pulsar}. The value of C$_{pix}$ scales with the pixel area and can be 20-150~fF. Thereby, $\tau$ can range between a few 10~ns to a few $\mu$s. It has to be kept in mind that polysilicon resistors are temperature dependent, and so is the pixel recovery time. Alternative quenching methods are being tested as well like integrated FET circuits in the silicon wafer~\cite{Simpl} or CMOS quenching circuits in the case of digital SiPMs~\cite{dSiPM}. 

\subsection{ Gain}

The SiPM gain, M is defined as the number of electric charges generated by one Geiger discharge in a SiPM pixel. This value is quite stable for Geiger discharges, and can be around 10$^5$-10$^7$. This is why a SiPM single pixel signal can be seen directly on the oscilloscope without preamplification as a signal of a few mV on a load of 50~$\Omega$. 
The pixel recovery time affects the design of the readout electronics as the integration or shaping time needed to exploit the large SiPM gain depends on the length of the signal.
The effective SiPM gain, M$_{eff}(\Delta t)$ is the number of electric charges measured in a given time interval after the Geiger discharge of one pixel, M$_{eff}(\Delta t) \le $ M. When reading the SiPM signal through a small decoupling capacitor or with a short shaping time, it is sometimes necessary to use a preamplification factor of 10-50. More discussion on the interplay between SiPM signal and design of the readout electronics is provided in section \ref{sec:chips}.

\subsection{Dark rate and after-pulse}

\begin{figure}
\centering
\includegraphics[width=0.4\textwidth]{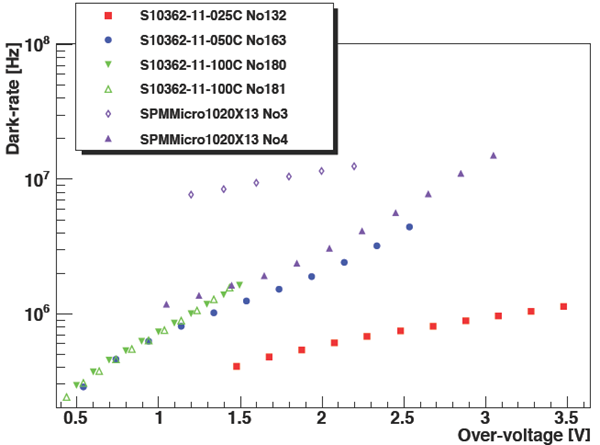}
\caption{\label{fig:dr}\em Dark rate comparison for various SiPMs from Hamamatsu (extension S10XX) and from SensL (extension SPMMicroXX). At the same over-voltage SiPMs with smaller pixel size have a smaller dark rate. In the Hamamatsu naming scheme ``S10362-11-XXXC'',  XXX corresponds to the pixel size in $\mu$m, while the two SensL devices have both a pixel size of 20$\times$20\,$\mu$m$^2$. Plot from~\cite{Tadday}.}
\end{figure}
In the absence of light a signal in the SiPM can be produced by carriers thermally generated in the depletion region.  
Secondary avalanches, the so called after-pulses, can follow a primary one caused by the release of carriers trapped in the silicon crystal during the primary Geiger avalanche. 
The first contribution is temperature dependent and approximately doubles for an increase of eight degrees. The second one depends on the pixel recovery time, the shorter $\tau$ is (20\% for Hamamatsu MPPCs~\cite{MPPC} with $\tau \le$ 30~ns, 3\% for CPTA/Photonique~\cite{CPTA} with $\tau \ge$ 150~ns,~\cite{Musienko}) the larger is the contribution from after-pulse. It also increases exponentially with the over-voltage\footnote{It has become SiPM jargon to refer to the operation voltage point above breakdown as {\it over-voltage}. \\
The dark rate of SiPMs can largely vary between 10$^5$-10$^7$~Hz. As shown (figure~\ref{fig:dr}) in a characterization study reported in~\cite{Tadday}, the dark rate depends exponentially on the voltage above breakdown at which the device is operated (higher dark rate for higher over-voltage, or $\Delta U = U - U_{bd}$. For simplicity I will use this abbreviation in this paper.}) and on the pixel capacitance (higher dark rate for larger capacitance at the same over-voltage). In addition, the dark rate is strongly influenced by the quality of the silicon wafer used for the SiPM production and by the level of cleanliness during the implant processes.  

\subsection{Optical crosstalk}

\begin{figure}
\centering
\includegraphics[width=0.4\textwidth]{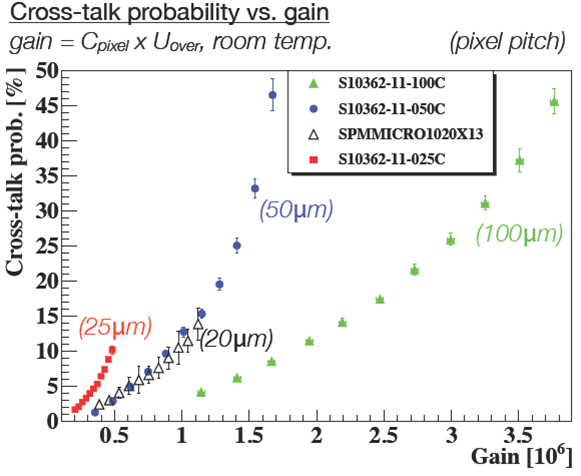}
\caption{\label{fig:ct}\em Crosstalk comparison for various SiPMs from Hamamatsu (extension S10XX) and from SensL (extension SPMMicroXX). At the same over-voltage SiPM with smaller pixel size have larger optical crosstalk probability. In the SensL device tranches are implemented which reduce crosstalk significantly. Plot from~\cite{Tadday}.}
\end{figure}
During the avalanche process in a p-n junction photons in the visible range are emitted. A typical number if of about 3$\cdot$10$^{-5}$ photons per charge carrier with a wavelength less than 1\,$\mu$m~\cite{Lacaita}. If these photons reach a neighboring pixel an additional avalanche can be caused. This effect is known as optical crosstalk between pixels and can occur as often as 30-40\% of the times a Geiger discharge takes place. The optical crosstalk increases exponentially with the SiPM gain. This is the combined effect of two factors: The number of photons generated in the discharge increases with the number of carriers, and the avalange trigger efficiency increases with increasing over-voltage.
Photons traveling directly between pixels can be stopped by optical tranches in the silicon.  
This technique introduced first by CPTA/Photonique~\cite{CPTA_tranches}, and later on used by SensL~\cite{SensL} and STMicroelectronics~\cite{STM}, reduces significantly the optical crosstalk to $<$1-3\%. The remaining effect is due to absorbed photons in the silicon bulk, much below the optical tranch, which are later re-emitted at an angle that allows them to reach a neighboring pixel. This last effect can hardly be avoided. A comparison study from~\cite{Tadday} of crosstalk of various devices is reported in figure~\ref{fig:ct}.\\

\section{SiPM response}

\noindent The SiPM response is the sum of the charges from all the pixels fired by a given amount of photons impinging on the SiPM active area. For very low number of impinging photons, N, the number of pixels fired, $N_{pix}$ is given by the product of N and the photon detection efficiency (PDE). $N_{pix}$ is strongly non-linear, and ignoring recovery time effects it saturates to the total number of pixels in a device for very large number of impinging photons. 

\subsection{Photon detection efficiency}

The photon detection efficiency can be factorized in three components:
\begin{equation}
  \label{eq:PDE}
  PDE(\lambda,U,T) = QE(\lambda) \cdot G_{ff} \cdot P_{bd}(\lambda,U,T),
\end{equation}
where $QE(\lambda)$ is the wavelength-dependent quantum efficiency of a pixel, $G_{ff}$ is the geometric fill factor of the SiPM, and  $P_{bd}(\lambda,U,T)$ is the wavelength-, voltage- and temperature-dependent probability for a photoelectron to trigger a breakdown, or Geiger discharge. 

\begin{figure}
\centering
\includegraphics[width=0.2\textwidth]{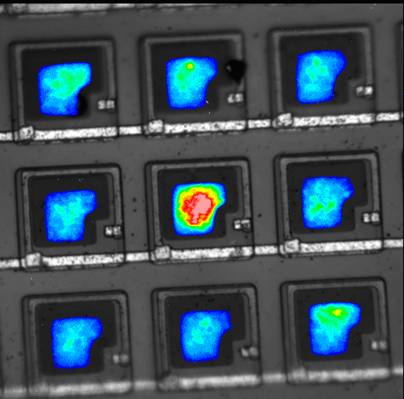}
\includegraphics[width=0.2\textwidth]{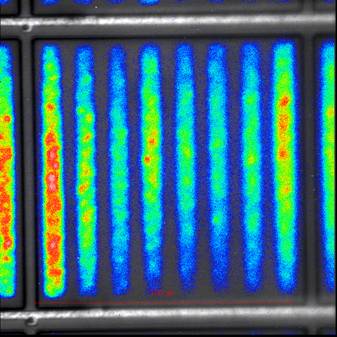}
\includegraphics[width=0.2\textwidth]{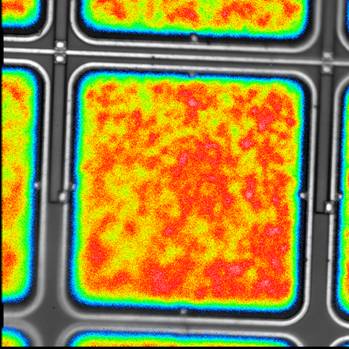}
\caption{\label{fig:ff}\em Photo-emission spectra for SiPM from different producers. Top left is a MEPhI/PULSAR, top right a SensL and bottom is a MPPC device. Aside of the pixel geometrical size there is an additional contribution to the fill factor from the dopant implant in the silicon. Plots from~\cite{Jelena}}
\end{figure}

The fill factor depends inversely on the number of pixels per unit area, and on the quenching technology used. Also the process used to dope the silicon can have an effect on the fill factor. Examples of photo-emission spectra for SiPM from different producers are given in figure~\ref{fig:ff}. The colored area in the pictures correspond to the active areas in the pixel and the color scale gives an impression of the electric field uniformity across the pixel. 
While the oldest devices have $G_{ff}$ = 20-30\%, with recent developments these values have increased to $\sim$70\% for CPTA/Photonique and $\sim$50\% for FBK-irst~\cite{FBK} SiPMs.

Concerning the wavelength sensitivity, the standard SiPM structure from most of the producers (CPTA/Photonique, SensL, FBK-irst, MEPhI/PULSAR~\cite{Pulsar}) is an n-doped implant on a p-type substrate. In this structure the thickness of the depleted region ranges from 3 to 10~$\mu$m and the depth of the n$^+$-p junction ranges from 0.3 to 1~$\mu$m~\cite{buzhan}. This structure is optimal for green-red light detection (1-3~$\mu$m absorption depth in silicon). 
To enhance the sensitivity to blue/UV light (400-450\,nm) some producers (Hamamatsu, and recently  MEPhI/PULSAR and STMicroelectronics) have opted for an inverted structure with a p-doped implant on an n-type substrate. In this case the blue/UV photons (absorption depth in silicon 100-500\,nm) convert in the p$^+$ region close to the surface. In the p$^+$ region the produced electrons drift towards the multiplication region, or p$^+$-n junction, while the holes drift towards the surface of the photodetector, or cathode. Electrons have a larger probability to trigger a breakdown in silicon than holes due to their larger ionization coefficient. A conversion in the p$^+$ layer has the highest probability to start a breakdown~\cite{oldham}.                                                     

\begin{figure}[!t]
\centering
\includegraphics[width=0.4\textwidth]{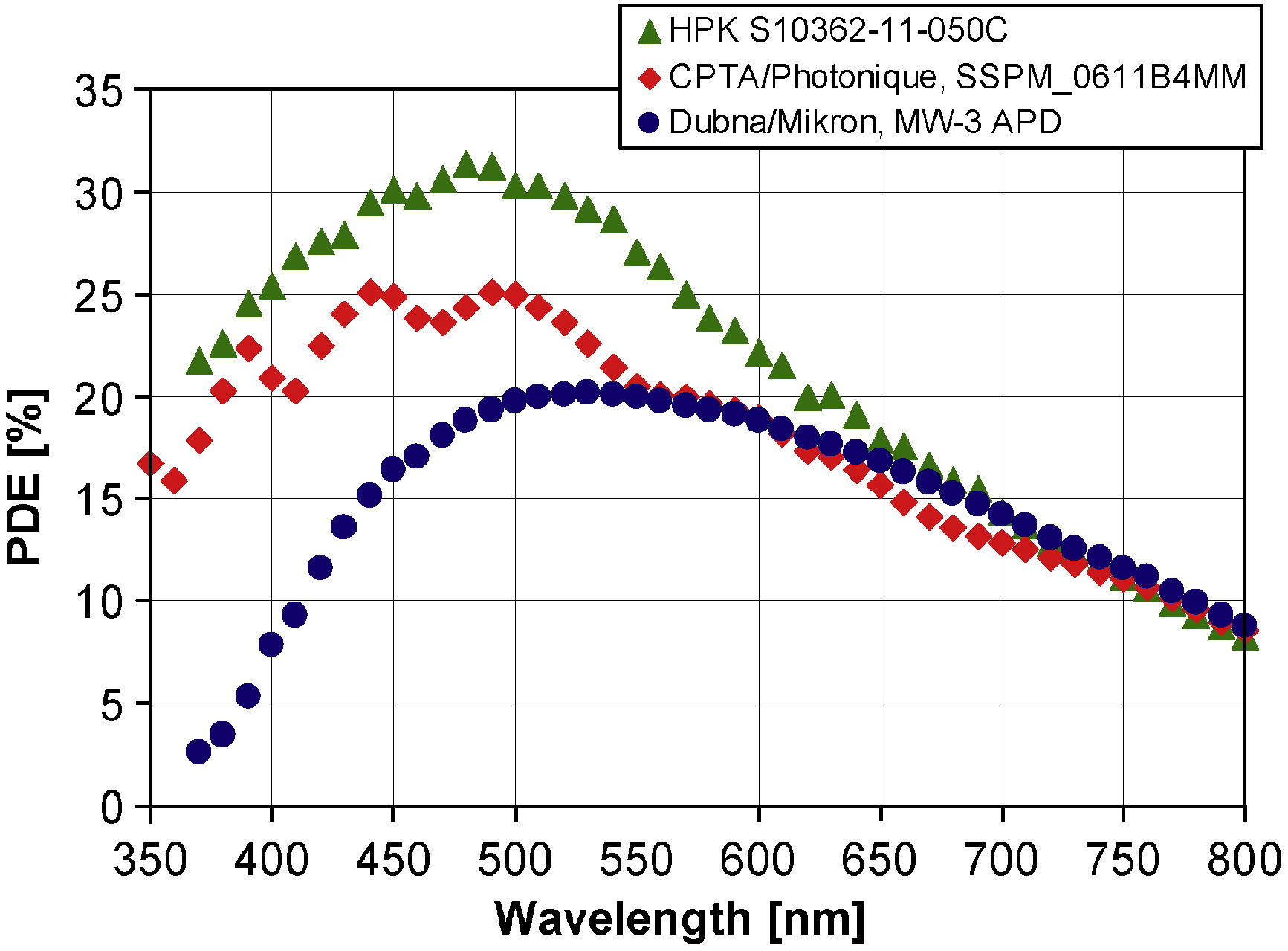}
\includegraphics[width=0.4\textwidth]{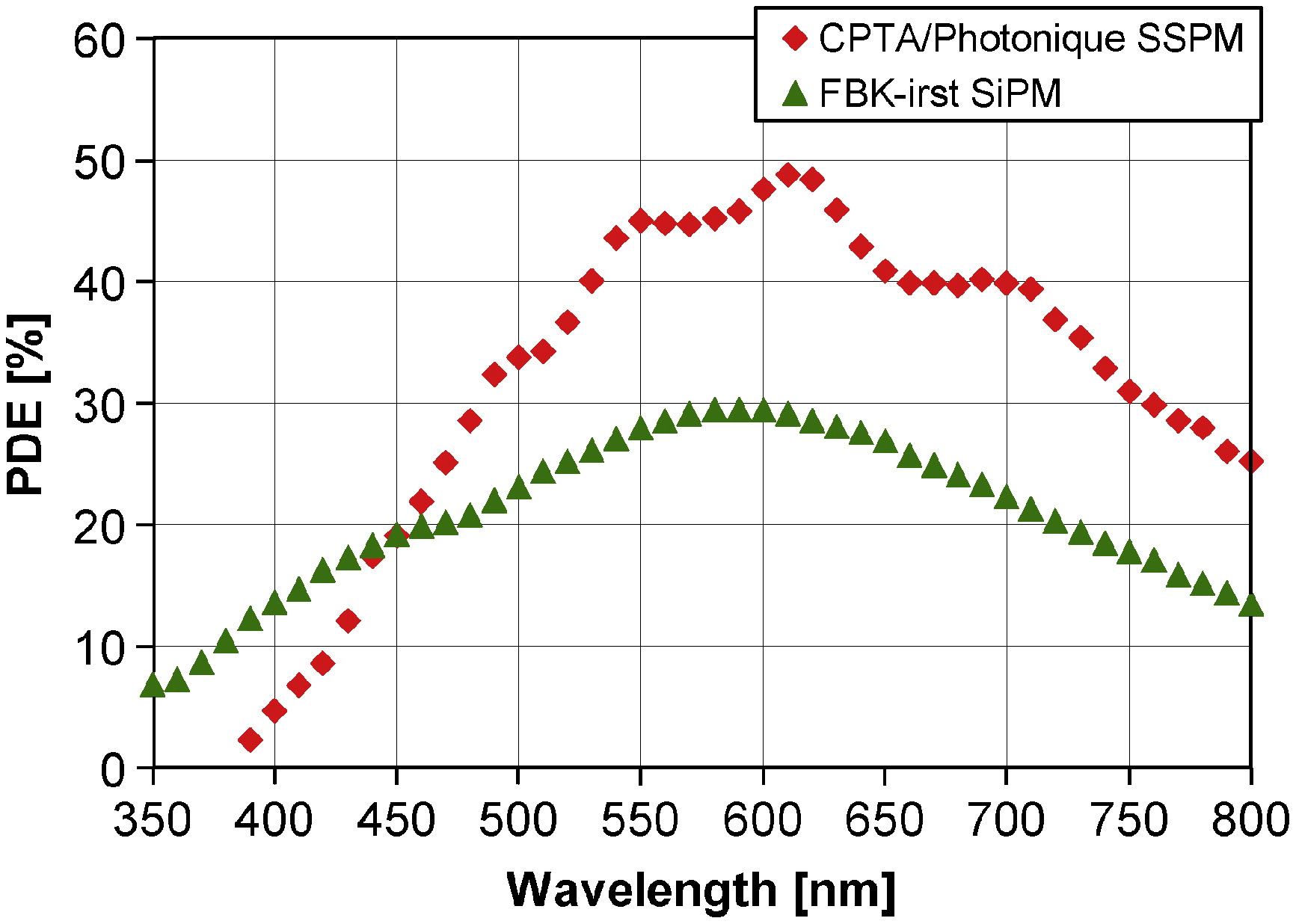}
\caption{\label{fig:pde}\em Photon detection efficiency as a function of the wavelength of light for n-type Hamamatsu (U=70.3 V), CPTA/Photonique (U=36 V) and Dubna/Mikron (U=119.6 V) G-APDs, measured at room temperature (top). Photon detection efficiency as a function of the wavelength of light for p-type CPTA/Photonique (U=42 V) and FBK-irst (U=35.5 V) G-APDs, measured at room temperature (bottom). Plots from~\cite{Musienko}.}
\end{figure}

Recent comparison studies of PDE from various SiPMs have been performed for example in~\cite{Tadday, Musienko}. Figure~\ref{fig:pde} reports results from~\cite{Musienko}. These measurements are properly subtracting the influence of optical-crosstalk and after-pulse, which if not accounted for leads to an enhanced PDE. Currently the best PDE in the blue/UV range is provided by the Hamamatsu MPPC (PDE$_{400-450\,nm}$~$\sim$35\% for devices with 100~$\mu$m pixel size, and  PDE$_{400-450\,nm}$~$\sim$30\% for devices with pixels of 50~$\mu$m size). A PDE$_{500-600\,nm}$~$\sim$40-45\% has been obtained by the n-on-p devices from CPTA/Photonique.

\subsection{Response function and dynamic range}

\begin{figure}
\centering
\includegraphics[width=0.4\textwidth]{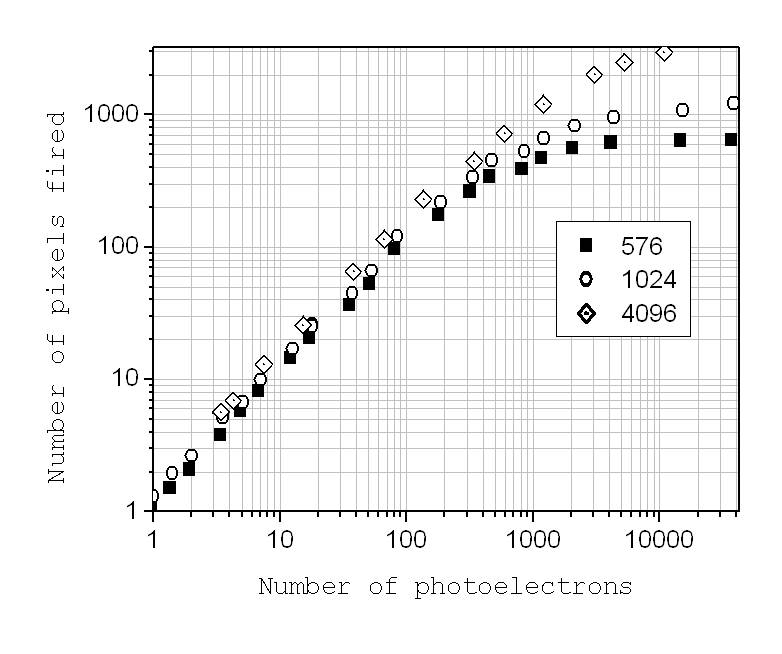}
\caption{\label{fig:sat}\em Non-linear response of MEPhI/PULSAR SiPMs with different number of pixels. The light signal is produced by a fast laser (40~ps). Plot from~\cite{BorisD}.}
\end{figure}
Due to the limited number of pixels and the finite pixel recovery
time, the SiPM is an intrinsically non-linear device. 
The response function of a SiPM correlates the observed number of pixels 
fired, $N_{pix}$,  to the effective number of photo-electrons generated, $N_{pe}$, including crosstalk and after-pulses.
The response of a SiPM can be approximated by the function 
 \begin{equation}
  \label{eq:sat}
  N_{ pix} = N_{ tot}\cdot\left(1-e^{-N_{ pe}/N_{ tot}}\right),
\end{equation}
  with $N_{ tot}$ being the maximum number of fired pixels in a SiPM.
This formula is a useful approximation for the case of uniform light
distribution over the pixels and short light pulses compared to the 
pixel recovery time.  
Signals equivalent to 80\% of the total number of pixels in a SiPM are 
underestimated by 50\% with respect to the measurement of a linear device.
This effect limits the dynamic range of a system
and reduces the acceptable spread in light yield of the system components.    
In addition, it requires a correction of the non-linear response which can 
be either an individual or a global curve, depending on how large the spread in PDE and crosstalk 
is of the various SiPMs in the system.
Measurements of the SiPM response curve for MEPhI/PULSAR devices with various number of pixels
are presented in figure~\ref{fig:sat}.

A significant improvement of the SiPM dynamic range is provided by the 
innovative design of the MAPD from Zecotek~\cite{Zecotek}. 
This is a non-conventional SiPM in the sense that it consists of 
a double n-p-n-p junction 
with micro-well structures located at a depth of 2-3~$\mu$m below the surface. 
The multiplication regions are just in front of the wells. This technique allows  
the quenching of the discharge in the absence of an additional resistor. In this way 
a pixel density of 10000-40000~mm$^{-2}$ is possible on an area up to 
3$\times$3~mm$^2$. 

\section{SiPM application to HEP detectors}
\label{sec:application}
One of the main advantages of SiPMs in the application for HEP detectors 
is the possibility to significantly increase the detector granularity 
with respect to PMT- and even APD-based readout designs.
Plastic scintillator with wavelength shifter (WLS) readout is a very attractive 
solution for detectors which require high granularity. Extreme examples are calorimeters 
for particle flow applications, as those designed for the future linear collider detectors~\cite{ILD}. 
Such calorimeters require single cells with a size of 3.0$\times$3.0~cm$^2$  and even 1.0$\times$4.5~cm$^2$.
Furthermore, the calorimeters are inside the spectrometer magnet so that 
the photodetectors used in the readout must be magnetic field insensitive. 
About five to ten million single cells are foreseen for a hadronic calorimeter barrel detector, 
similar numbers are needed also in the end cap detectors. For the electromagnetic calorimeter the situation is 
similar if one chooses to use the scintillator technology for the readout~\cite{ScECAL}. 
But also more modest projects require thousands of SiPMs, the T2K detectors~\cite{T2K} use about 56000 MPPCs.
These extraordinary numbers are only achievable with a photo-sensor technology which is robust, simple and cheap. 
In addition, they require to establish a solid mass production chain 
and to define clear selection requirements on the basic SiPM parameters.

\subsection{SiPM mass production}
\label{sec:production}

While more producers are developing towards mass production possibility, at present 
we have experience with mass tests only for few experiments.

\begin{figure}
\centering
\includegraphics[width=0.23\textwidth]{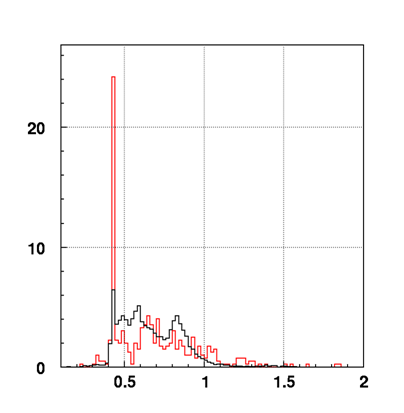}
\includegraphics[width=0.23\textwidth]{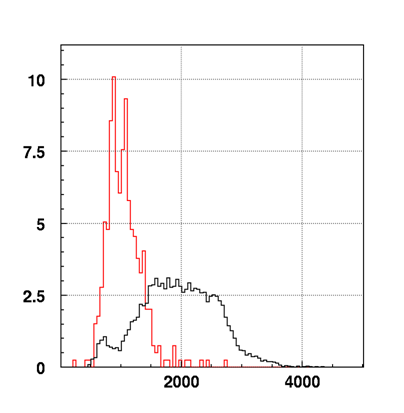}
\caption{\label{fig:spread}\em Spread of SiPM parameters from the characterization study of 8000 MEPhI/PULSAR (black histogram) and 1000 CPTA/Photonique SiPMs (red histogram) for the construction of the AHCAL CALICE prototypes~\cite{HCAL,EUDET}. Left) SiPM gain in units of [10$^6$]. Right) The dark-rate at 0.5 p.e. in units of [kHz].}
\end{figure}

The very first detector built and operated with about 8000 SiPMs from MEPhI/PULSAR 
is the CALICE hadronic calorimeter prototype for an ILC detector~\cite{HCAL,EUDET}.
An example of the characterization studies and selection criteria applied on the 
MEPhI/PULSAR SiPMs is reported in figure~\ref{fig:spread}. 
SiPMs are selected with a noise rate of less than 3~kHz at 7.5 pixels threshold, 
corresponding in this detector to half of the signal of a minimum ionizing particle. Additional
requirements on optical crosstalk and SiPM current reduce the yield only slightly. 
The response and the noise of the SiPM are sensitive to temperature variations.
A decrease by one degree leads to a decrease of the breakdown
voltage by 50~mV, which is equivalent to an increase of the bias voltage by the same amount.
Being the very first mass produced SiPMs, these devices came from various, quite different
batches. The operation voltage of each batch has a spread of about 5~V, and the SiPM gains are in the 
range 0.3-1.3$\cdot 10^6$.

The spread of parameters has consequences for the design of 
the readout electronics. 
Each SiPM requires individual voltage adjustment with a precision better then 50~mV, and 
a stability of the supplied power of better then 10~mV. 
To better cope with the spread in SiPM gain it is desirable to have 
individual preamplification adjustment for each channel in the 
readout chip, or individual adjustable zero suppression thresholds. 
Examples of readout chip architectures are discussed in section~\ref{sec:chips}.

The second generation prototype~\cite{EUDET} of the CALICE hadronic calorimeter,
is being equipped with CPTA/Photonique SiPMs of the type n-on-p. 
The test of the first thousand pieces produced shows a significant 
reduction of dark rate and noise compared to the earlier SiPMs\footnote{This is not intended as a comparison of the quality of the two producers, as significant improvement has come from the first experience of MEPhI/PULSAR.}. The spread in SiPM gain and in operation 
voltage remains similar, and with that the implications on the design of the readout 
electronics. 

From the characterization studies of the Hamamatsu MPPC used in the detector for T2K~\cite{T2K_MPPC} 
one sees the range of operation voltages is reduced to 3\,V, but still too large to 
remove the requirement of individual voltage adjustment. The SiPM gains for about 
17000 devices 
are in the range 4-5.5$\cdot 10^5$ and the dark rate at 0.5 pixel threshold ranges 
between 300~kHz and 800~kHz. These SiPMs were selected with a failure rate of less 
then 0.05\%.

\subsection{Application of SiPMs in calorimetry}
\begin{figure}
\centering
\includegraphics[width=0.3\textwidth]{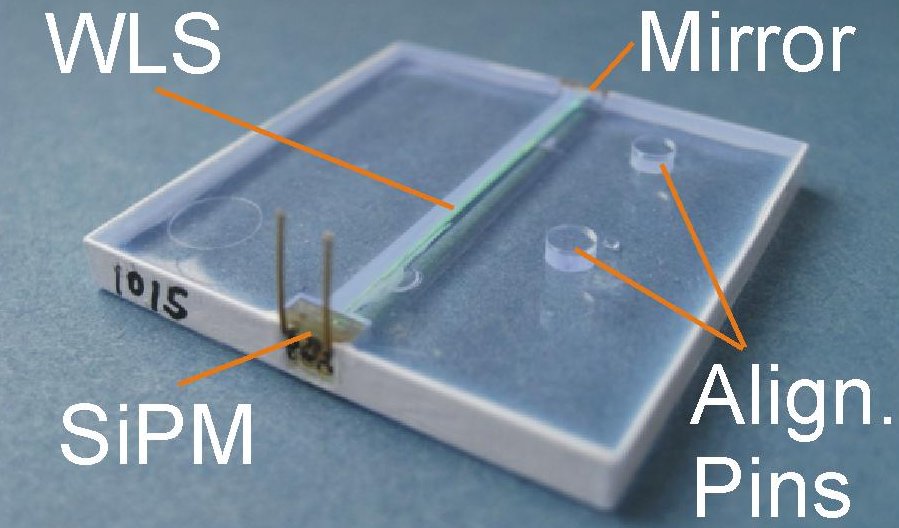}
\includegraphics[width=0.3\textwidth]{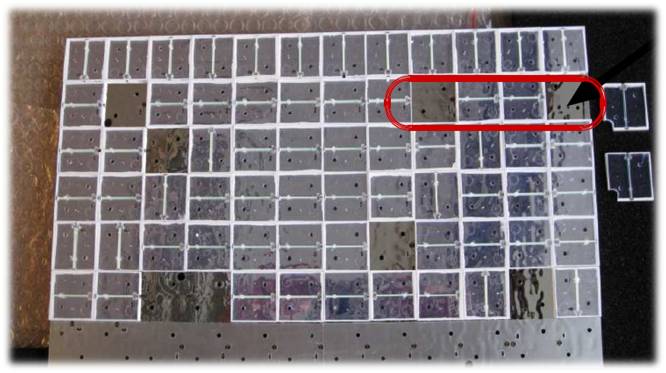}
\caption{\label{fig:HCAL}\em Top) Single cell of the CALICE AHCAL calorimeter described in~\cite{EUDET}: A scintillating tile with embedded wavelength shifting fiber, SiPM, mirror and alignment pins. Bottom) One calorimeter module hosting 36$\times$36 single scintillator tiles on the back side of the readout electronics.}
\end{figure}

\begin{figure}
\centering
\includegraphics[width=0.49\textwidth]{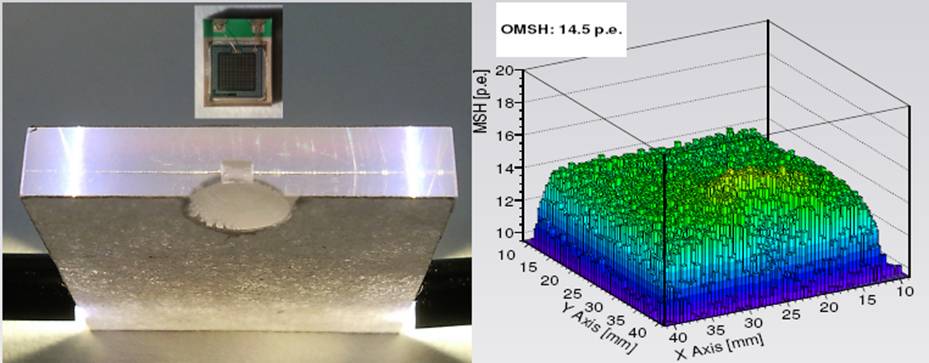}
\caption{\label{fig:MPItile}\em Left) Special shaped tile of 5 mm thickness with a depression and a slit for the integration of a surface mount MPPC25-P. Right) Signal amplitude measured over the surface area, showing a high degree of uniformity. The
mean amplitude is 14.5 p.e, \cite{Simon}.}
\end{figure}

As already mentioned the pioneering detector R\&D for the application of SiPMs in calorimetry 
is the analog hadronic calorimeter (AHCAL) of CALICE~\cite{HCAL,EUDET}. 
The most recent design for the scintillator tile/SiPM system for this project is shown in figure~\ref{fig:HCAL}. A plastic scintillator tile (top) of 3.0$\times$3.0$\times$0.3~cm$^3$ is read out via a CPTA/Photonique SiPM (796 pixels). 
The scintillation light is shifted to the green wavelength and coupled to the photo-detector via a WLS fiber with a mirror at the opposite side of the SiPM. 
About 150 of these tiles are assembled on the back of one PCB carrying the readout electronics. This constitutes one 
calorimeter module (figure~\ref{fig:HCAL} bottom). 
The alignment of the tiles is ensured by pins realized during the molding of the scintillators. 
For each calorimeter module the analog signals from the SiPMs are read out by four 36-channel ASICs, the SPIROC chip. With this design a 
calorimeter cell offers a detection efficiency for single minimum ionizing particles of 95\% with noise hit probability of 10$^{-4}$ above threshold. 
An optimized design of the calorimeter basic unit, the tile, is proposed in~\cite{Simon} (see figure~\ref{fig:MPItile}), where blue-sensitive SiPMs are directly coupled to plastic scintillators without a WLS fiber. To achieve satisfactory uniformity of response over the tile surface a special shaping of the coupling position has been developed. The new tile geometry is well suited for mass production and shows good overall performance in terms of uniformity and light yield.\\
Alternative to the tile shape, mini-strips are a promising solution to increase the effective granularity of plastic scintillator calorimeters even further. The smallest cell size of this kind is designed for the CALICE electromagnetic Tungsten-scintillator sampling calorimeter (ScECAL,~\cite{ScECAL}). Scintillator strips of size 1.0$\times$4.5$\times$0.3~cm$^3$ are manufactured by an extrusion method with a central hole of 1\,mm diameter to place the WLS fiber.  
Tests have been made with a co-extruded coating of TiO$_2$ (for light shielding), but the uniformity of the light collection along the strip was not satisfactory. The final design adopts reflector foils to optically isolate each individual strip in a detector layer~\cite{ScECAL2}.
A similar strip design has been used extensively in several detectors for the T2K experiment. 

\begin{figure}
\centering
\includegraphics[width=0.35\textwidth]{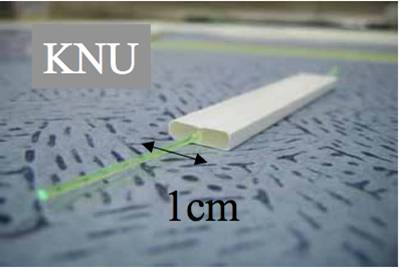}
\caption{\label{fig:ScECAL}\em Extruded scintillator strip with TiO$_2$ coating and 1~mm diameter WLS fiber, produced for the CALICE ScECAL~\cite{ScECAL2}.}
\end{figure}

\subsection{Calibration and monitoring system}
One of the advantages of using SiPMs in a system is the possibility to calibrate and monitor the gain of the device from the single photoelectron peak spectrum. This requires illuminating each SiPM with low light intensity (1-2 p.e.). As the SiPM gain is sensitive to temperature and voltage changes, monitoring the gain allows to correct for variations in the detector response. The advantage of this method is that it is insensitive to small variations in the light intensity emitted from the light source. 
Alternatively, the response of the SiPM can be monitored using a medium light intensity (20-100 p.e.), but in this case the stability of the light intensity has to be ensured either by construction (laser) or via extra monitoring (LED+PIN photo-diode system). For light intensities larger than a few 100 p.e. the effect of the SiPM non-linearity becomes relevant. \\
Monitoring the saturation point of the SiPM response function allows to monitor the number of active pixels in the device. This is possibly relevant in case of expected aging due to operation at very high over-voltage or in a hard radiation environment. In this case the light monitoring system has to be capable of delivering a high light intensity, typically a factor 2-3 larger than the number of pixels in the SiPM. Furthermore, the duration of light emission must be small (typically 10 ns) compared to the emission spectrum of the scintillator system. Such requirements are not easy to meet for thousands of channels. Two alternative design concepts are presented in~\cite{IEEE_peter}, one based on a central driver and optical signal distribution via notched clear fibers, one based on electrical signal distribution and an individual SMD-type LED per calorimeter cell. With both systems single photoelectron peak spectra are nicely obtained, while studies are still ongoing to demonstrate if the SiPM saturation can be monitored.

\begin{figure}
\centering
\includegraphics[width=0.35\textwidth]{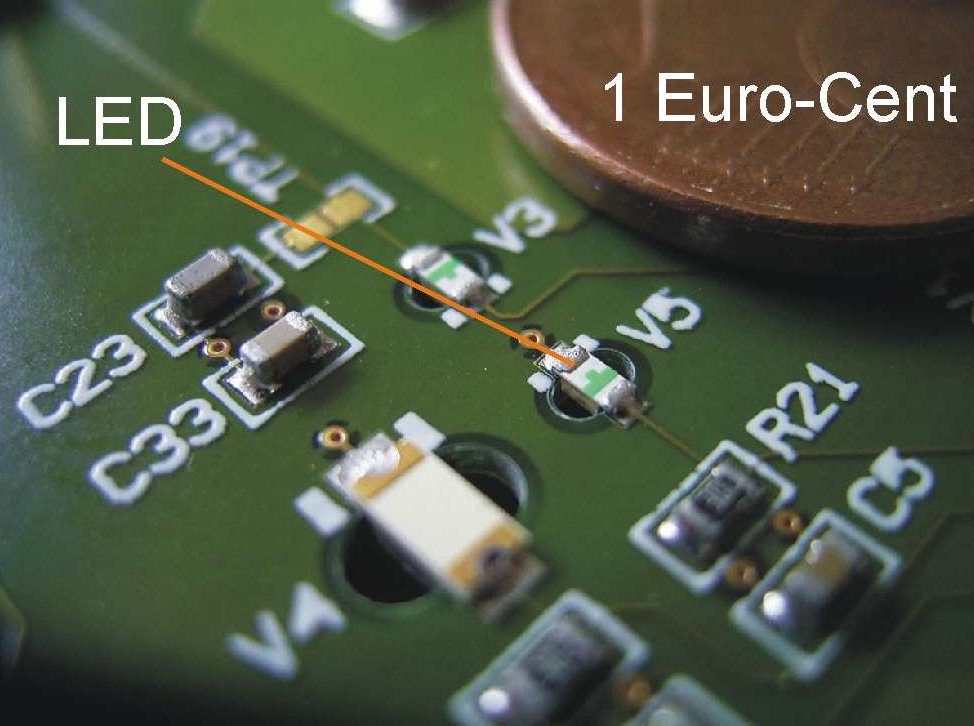}
\caption{\label{fig:LED}\em 
One option for the AHCAL LED system based on electrical signal distribution and an individual SMD-LED per tile~\cite{IEEE_peter}.}
\end{figure}

\begin{table*}[!h]
  \begin{tabular}{|l|c|c|c|c|}
\hline
Chip name & Measured quantity & Application & Input configuration & Technology \\
\hline
FLC\_SiPM & Pulse charge & ILC analog HCAL & Current input & CMOS 0.8um        \\
MAROC2    & Pulse charge, trigger & ATLAS lumi & Current input & SiGe 0.35um        \\
SPIROC    & Pulse charge, trigger & ILC HCAL & Current input & SiGe 0.35um        \\
NINO      & Pulse width, trigger & ALICE ToF & Differential input & CMOS 0.25um        \\
PETA      & Pulse charge, trigger, time & PET & Differential input & CMOS 0.18um        \\
BASIC     & Pulse height, trigger & PET & Current input & CMOS 0.35um        \\
SPIDER    & Pulse height, trigger, time & SPIDER RICH & Current input &         \\
RAPSODI   & Pulse height, trigger & SNOOPER & Current input & CMOS 0.35um        \\
\hline
  \end{tabular}

\vspace{0.3cm}

  \begin{tabular}{|l|c|c|c|c|c|c|}
\hline
Chip name & N of chan. & Digital  & Area [mm$^2$] & Dynamic  &Input R [$\Omega$] & Timing \\
          &            &  output  &               &  range   &                   & jitter [ps] \\
\hline
FLC\_SiPM & 18 & n& 10 &        &    &   \\
MAROC2    & 64 & y& 16 & 80 pC  &50  &   \\
SPIROC    & 36 & y& 32 & 2000 pe&    &   \\
NINO      & 8  & n& 8  & 2000 pe&20  &15 \\
PETA      & 40 & y& 25 & 8 bit  &    &50 \\
BASIC     & 32 & y& 7  & 70 pC  &17  &120\\ 
SPIDER    & 64 & n& 15 & 12 pC  &    &   \\
RAPSODI   & 2  & y& 9  & 100 pC &20  &   \\
\hline
  \end{tabular}
  \caption{List of existing SiPM dedicated readout chips and their properties, from W. Kucewicz in~\cite{Kucewicz}. }
  \label{tab:SIPMchip}
\end{table*}

\section{SiPM dedicated readout chips}
\label{sec:chips}
SiPMs are an excellent device for highly granular multi-channel systems. To exploit at best the advantages of SiPMs they need to be read out via a dedicated multi-channel chip. 
Depending on the specific application few chips have established themselves among the users community. 
Table~\ref{tab:SIPMchip} shows an overview edited by W. Kucewicz in~\cite{Kucewicz}. A comprehensive list of 
references to individual publications for each chip can be found in this document.

The table compares eight SiPM-dedicated ASIC chips developed for applications in HEP, medical 
physics and homeland security. The chips can be distinguished in two categories according to 
the measurements they allow to perform on a given signal from the detector: only energy (either 
through a charge measurement, or a pulse height / pulse width measurement), or energy and time.
Furthermore, they have to be distinguished for the output they offer, analog or digital. An already digitized output 
has the clear advantage of a simpler and cheaper readout electronics stage after the chip.\\
For instance, the main new features of the SPIROC compared to its predecessor, the FLC\_SiPM chip 
are the integration of the digitization step (12-bit ADC and 12-bit TDC for charge 
and time measurements) and the self-triggering capability with an adjustable zero suppression threshold, 
which acts as an on-detector zero suppression. \\

\begin{figure}
\centering
\includegraphics[width=0.49\textwidth]{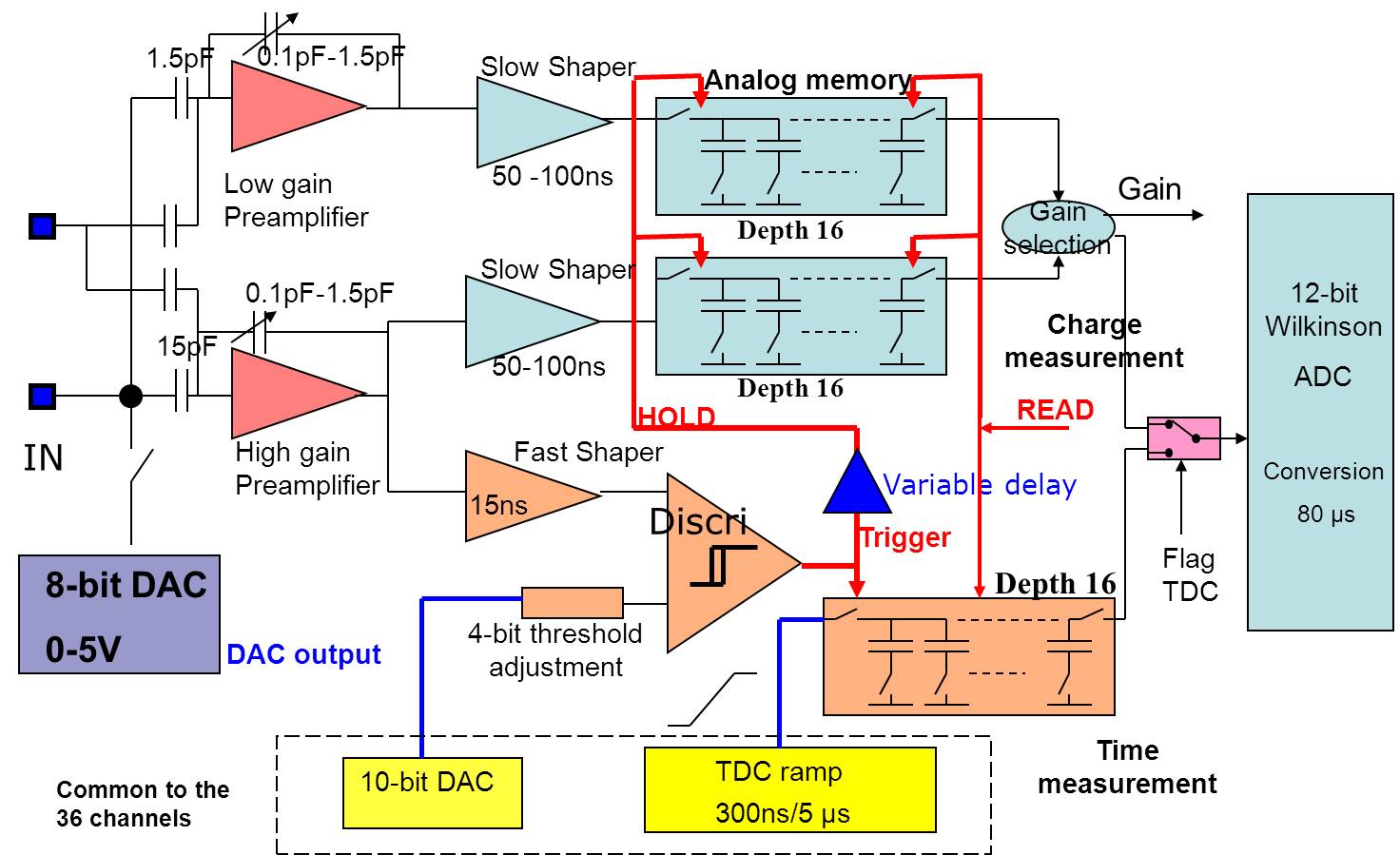}
\caption{\label{fig:SPIROC}\em Schematic circuit of one channel of the SPIROC chip~\cite{SPIROC}). The SiPM is connected at the input IN to the preamplifier via a coupling capacitor.}
\end{figure}

When working with a large number of SiPMs not preselected in terms of operation voltage
it is necessary to adjust each bias voltage individually.
This is easiest achieved with a programmable DC voltage level added to the SiPM bias line.
The SPIROC, FLC\_SiPM and SPIDER chips for example provide a DAC-steerable voltage of typically 5~V on each input line for a channel-wise adjustment.
An example is shown in figure~\ref{fig:SPIROC}, where 
the SiPM anode is connected directly to the chip input pin (IN) using an RC filter to ground (not shown) with R=50~$\Omega$ and C=100~nF. The voltage generated on the 50~$\Omega$  resistor is amplified by an AC-coupled low noise charge sensitive amplifier. The combination of this coupling capacitance  and the charge sensitive amplifier results in a voltage readout scheme. In order to fit inside the chip package the coupling capacitor cannot be too large. In this design C=1.5~pF is used for the low gain line and C=15~pF for the high gain line. The effect is that only a small portion of the about 10$^6$ electrons generated in one Geiger discharge are effectively measured. For a SiPM gain smaller than 5$\cdot$10$^5$ the single photoelectron peak spectra cannot be properly resolved with this method. 

Recently new SiPMs with gains of 2.5$\cdot$10$^5$ and lower have become available. Hence, a charge sensitive readout scheme is being investigated which could make the SPIROC chip even more versatile in dealing with a large variety of SiPMs. One solution is the use of a current conveyor before preamplification at the same time removing the 50\,$\Omega$ termination resistor in the voltage scheme. In this approach the bias-tuning DAC voltage is coupled onto the signal line, while the input current is scaled down and sent to the preamplifier. A similar circuit is implemented in the BASIC chip to fine tune the bias voltage for each of the 32 channels.    

A variable gain in the chip preamplifier stage compensates the SiPM gain dispersion.
The first three chips in the list have programmable preamplifiers, which allow to measure input signals in the range from 1 to 2000 p.e.

For some specific applications, like operation in the dense environment of a sandwich calorimeter at 
the ILC, ultra-low power consumption is an important issue. 
The chips in the ``ROC'' family are designed to operate with pulsed power supply for minimized heat dissipation. 
The foreseen power consumption amounts to 25\,$\mu$W per channel for the final ILC operation. 

\section{Conclusions}
Silicon Photomultipliers are a novel, yet well established technology for detection of visible photons. 
They are a very promising alternative to standard photo-multiplier tubes in many applications in high energy 
physics and also in medical fields. 
Their small size makes them suitable for the direct single-channel readout of highly segmented detectors, 
like for example calorimeters, fiber tracking detectors and positron emission tomography detectors.
Though the application of SiPMs is rapidly growing and extending to always new designs the mass production capability 
of these photo-detectors is still not fully proportionate to the request. The physics community recognizes the need of more 
competitive products commercially available in large quantities and with a minimum spread of the key parameters like operation voltage, 
gain, dark rate, crosstalk, photo-detection efficiency, etc..

For a detector operating a large number of SiPM a monitoring system is often mandatory. 
Two possible designs are discussed in which light from an LED is used to monitor the SiPM gain and 
response, and possibly the saturation point of the SiPM non-linear response function.

To fully exploit the advantages of SiPMs as photo-detectors, newly developed multi-channel readout chips are required. 
Available chips for SiPM readout are presented and theirs differences and advantages are discussed. To accommodate 
the large variety of existing SiPMs the ideal readout chip should offer single-channel adjustable voltage, 
preamplifier gain and/or zero suppression threshold.

\section{Acknowledgments}
I thank Yuri Musienko, Jelena Ninkovic, Wojtek Kucewicz, from whom I have taken large part 
of the material needed for this review. Furthermore, I thank my colloages from 
the CALICE collaboration for allowing me to present material and results from their research. 
Work supported by the Helmholtz-Nachwuchsgruppen grant VH-NG-206.

\end{document}